\documentclass[aps,prl,showpacs,nobibnotes,twocolumn,
nofootinbib,nobalancelastpage,superscriptaddress,preprintnumbers]{revtex4}
\usepackage{graphicx}
\usepackage{amsmath,color}
\usepackage{latexsym}
\usepackage{dcolumn}
\usepackage{bm}
\input{epsf}
\newcommand{\be}{\begin{equation}}
\newcommand{\ee}{\end{equation}}
\newcommand{\ba}{\begin{eqnarray}}
\newcommand{\ea}{\end{eqnarray}}
\newcommand{\bi}{\begin{itemize}}
\newcommand{\ei}{\end{itemize}}

\newcommand{\bfi}{\begin{figure}
\epsfxsize=9cm
\epsffile}
\newcommand{\efi}{\end{figure}}

\begin{document}

\title{Dark Matter Annihilation from Nearby Ultra-compact Micro Halos to Explain the Tentative Excess at $\sim$1.4 TeV in DAMPE data}
\author{Fengwei Yang}
 \altaffiliation{Physics Department, The University of Hong Kong}
 \email{fwyang@hku.hk}
 
\author{Meng Su}
\affiliation{Department of Physics and Laboratory for Space Research, The University of Hong Kong, PokFuLam, Hong Kong SAR, China}

\author{YUE ZHAO}
\affiliation{Tsung-Dao Lee Institute, and Department of Physics and Astronomy, Shanghai Jiao Tong University, Shanghai 200240}
\affiliation{Michigan Center for Theoretical Physics, University of Michigan, Ann Arbor, MI 48109}

\date{\today}

\begin{abstract}
The tentative 1.4 TeV excess in the $e^+e^-$ spectrum measured by The DArk Matter Particle Explorer (DAMPE) motivates the possible existence of one or more local dark matter concentrated regions. In particular, Ultra-compact Micro Halos (UCMHs) seeded by large density perturbations in the early universe, allocated within ~0.3 kpc from the solar system, could provide the potential source of electrons and positrons produced from dark matter annihilation, enough to explain the DAMPE signal. Here we consider a UCMH with density profile assuming radial in-fall and explore the preferred halo parameters to explain the 1.4 TeV "DAMPE excess". We find that typical parameter space of UCMHs can easily explain the ``DAMPE excess" with usual thermal-averaged annihilation cross section of WIMP. The fraction of dark matter stored in such UCMHs in the Galactic-scale halo can be reduced to as small as $O(10^{-5})$, well within the current cosmological and astrophysical constraints.
\end{abstract}

\maketitle
\section{Introduction}

The DArk Matter Particle Explorer (DAMPE; \cite{ChangJin:550, 2017APh....95....6C}) is the first Chinese space mission for astronomical study, it was successfully launched from Jiuquan Satellite Launch Center on December 17, 2015. It has already surveyed the full sky four times and collected almost two-year cosmic rays data continuously. Recently, the DAMPE collaboration has published the positron-electron energy spectrum as the first scientific results from DAMPE, which extends the direct detection of electron energy spectrum with a spaceborne detector to ~4.6 TeV \cite{Ambrosi:2017wek}.  

The DAMPE measurement has unprecedentedly high energy resolution, low particle background, and well controlled instrumental systematics. Although the majority of the spectrum can be fitted by a smoothly broken power-law model with a spectral break at E$\sim$0.9 TeV, a tentative peak at $\sim$1.4 TeV in $e^+e^-$ total spectrum has been claimed \cite{Ambrosi:2017wek}. The excess of the $e^+e^−$ pairs at 1.4 TeV is approximately  $2.5\times10^{-8}~\rm{GeV^{-1}~s^{-1}~sr^{-1}~m^{-2}}$ from the detected electron-positron energy spectrum of DAMPE \cite{Yuan:2017ysv}. Since the announcement of the results, there have been extensive discussion on the possible theoretical explanation and observational constraints on the explanations of the ``DAMPE excess'' with both particle origin or astrophysical origin     \cite{2017arXiv171203652O, 2017arXiv171203642S, 2017arXiv171203210Z, 2017arXiv171202939X, 2017arXiv171202744G, 
2017arXiv171202739T, 2017arXiv171202381L, 2017arXiv171202021D, 2017arXiv171201244C, 2017arXiv171201239G, 2017arXiv171201143Z, 2017arXiv171201032K, 2017arXiv171200941N, 2017arXiv171200922G, 2017arXiv171200869L, 2017arXiv171200793C, 2017arXiv171200372N, 2017arXiv171200370G, 2017arXiv171200362J, 2017arXiv171200037C, 2017arXiv171200011C, 2017arXiv171200005H, 2017arXiv171111579L, 2017arXiv171111563D, 2017arXiv171111452C, 2017arXiv171111376A, 2017arXiv171111333G, 2017arXiv171111182C, 2017arXiv171111058T, 2017arXiv171111052Z, 2017arXiv171111012D, 2017arXiv171111000G, 2017arXiv171110996F, 2017arXiv171110995F, 2017arXiv171110989Y}. 

The electron and positron excess appears in one energy bin around 1.4 TeV. Such sharp peak in $e^+e^-$ energy spectrum is unexpected because high energy electrons quickly lose energy through synchrotron radiation and Inverse Compton Scattering while propagating in the Milky Way. The sharp peak in the energy spectrum indicates a nearby source of the high energy electrons and positrons. If explained by DM, the source of such energetic and monoenergetic electrons and positrons has to be close to the solar system. 

The Ultra-compact Micro Halos (UCMHs) have been studied as a potential dark matter substructure at small scale. The cosmological and astrophysical constraints are widely discussed \cite{2012PhRvD..85l5027B, 2013PhRvD..87h3519Y, 2016arXiv161200169B}. Several mechanisms can easily induce UCMHs in the early universe, e.g. phase-transitions in the early universe, inflaton with large slop at late stage of inflation, or topological defects like cosmic strings \cite{2009ApJ...707..979R, 2012PhRvD..85l5027B}. UCMHs intrinsically have larger dark matter concentration and a steeper density profile than the predictions on DM subhalo from the standard cold DM simulations. They have higher chance to survive from tidal stripping in the Galaxy. Their existence may provide unique and strong DM annihilation/decay signals nearby solar system. Especially, one may expect large induced fluxes of cosmic rays from the nearby UCMHs, including high-energy positron-electron pairs, gamma rays, and neutrinos, which are much higher than those from the center of the Milky Way. On the other hand, these small mass UCMHs might not have accreted enough baryonic matter to provide observable signal in radio and gamma ray.

In this letter, we study the possibility of attributing the excess of the ``DAMPE excess'' to the DM annihilations in the vicinity of the solar system. We consider one or more UCMHs nearby as the cosmic ray source \cite{2000RPPh...63..793B, 2005PhR...405..279B, 2009NJPh...11j5006B}, where $e^+e^-$ flux is produced through the pair annihilation of dark matter particle. We explore if such scenario could explain the DAMPE signal with the limits on the abundance of these primordial UCMHs structures. We study the properties of the UCMH as a function of its distance to the solar system. We first describe the profile of UCMHs. With generic assumptions, we find the mass of UCMHs can be easily related to the effective radius. Then we provide more details on the $e^+e^-$ excess observed by DAMPE and we find the preferred parameters of UCMHs in order to explain the 1.4 TeV $e^+e^-$ excess.

\section{\label{sec:model}UCMHs' Model} 

There are many possible mechanisms to trigger the formation of UCMHs, such as sizable density fluctuations at small scales induced at the late stage of inflation.  An over-dense region can efficiently accumulate DM particles after matter-radiation equality. The accretion process will stop when dynamical friction becomes important. We take the cut off in accretion to be the beginning of star formation. Thus the current properties of UCMH nowadays is determined by the profile at $z\sim 10$. Note, that the UCMHs are dense and safe from tidal perturbations during the evolution \cite{Bringmann:2011ut}.

Such UCMHs, if exists, can induce interesting astrophysical signatures. The UCMHs are formed through radial in-fall \cite{2009NJPh...11j5027B,2016arXiv161200169B}, indicating the density profile of UCMH as
\begin{equation}
\rho_{UCMH}(z,r)=\frac{3f_{\chi}M_{UCMH}(z)}{16\pi R_{UCMH}^{3/4}(z)r^{9/4}},
\label{rho_zr}
\end{equation}
$f_{\chi}$ is the dark matter fraction in the patch when UMCHs are formed, $M_{UCMH}(z)$ is the mass of UCMH at redshift $z$, and $r$ is the distance away from the center of UCMH. The effective radius of UCMH,  $R_{UCMH}(z)$, is obtained from numerical simulations \cite{2008ApJ...680..829R,2007ApJ...662...53R}
\begin{equation}
R_{UCMH}(z)=0.019 \rm{pc}\left(\frac{1000}{1+z}\right)\left(\frac{M_{UCMH}(z)}{M_\odot}\right)^{1/3}.
\label{rz}
\end{equation}

The radial in-fall approximation breaks down when angular momentum of infalling gas becomes important. The scaling behavior, $r^{-9/2}$, in Eq. \ref{rho_zr} is truncated at a cut-off radius $r_c$ \cite{2012PhRvD..85l5027B},
\begin{eqnarray}
\frac{r_{c,ang}}{R_{UCMH}(0)}&=&\nonumber\\
2.9\times10^{-7}&\times&\left(\frac{1000}{1+z_{coll}}\right)^{2.43}\left(\frac{M_{UCMH}(0)}{M_\odot}\right)^{-0.06},
\end{eqnarray}
where $z_{coll}$ is the redshift when the UCMH collapsed, which we take to be around 1000 as the smallest allowed
redshift of collapse. The density is approximately constant for $r< r_c$. 

At the meanwhile, if DM annihilation cross section is sizable, such process is not negligible when DM density is high. This imposes an additional modification to the DM profile considered above, i.e. an upper limit on DM density,
\begin{equation}
\rho_{c,ann}(t)=\frac{m_{\chi}}{(t-t_i)\langle\sigma v\rangle},
\end{equation}
where $m_{\chi}$ is the mass of a dark matter particle. $t_i$ is taken to be the time of matter-radiation equlibrium, i.e. $t_i=t(z_{eq})=59 \rm{Myr}$. $t$ is the taken to be $13.799 \rm{Gyr}$ for UCMHs nearby. $\langle\sigma v\rangle$ is the thermally-averaged annihilation cross section of the dark matter particle. 

Thus the truncation of the UCMH profile is determined by the competition between $r_{c,ang}$ and $r_{c,ann}$, i.e. $r_c={\rm{max}}\left(r_{c,ann},r_{c,ang}\right)$. Take typical averaged-thermal annihilation cross section as a benchmark, $\langle\sigma v\rangle = 3\times 10^{-26}\rm{cm}^3/\rm{s}$, and set $m_\chi=1.4$ TeV, we find that as long as UCMH mass is larger than $1.67\times10^9$g, $r_c$ is always determined by $r_{c,ann}$.


The full piecewise expression of the density of an UCMH at some radius $r$ will be:
\begin{equation}
\rho_{UCMH}(r)=
\left\{
\begin{aligned}
\rho_c,&~0 \le r \le r_c\\
\rho_c (\frac{r}{r_c})^{-9/4},&~r_c < r \le R_{UCMH}(z)\\
0, &~r > R_{UCMH}(z)
\end{aligned}
\right.
\end{equation}
where $\rho_c$ is determined by the UCMH profile in Eq. \ref{rho_zr} at $r=r_c$. Further, since we are focused on UCMHs nearby the solar system, these UCMHs should have already stopped accretion process, and $z$ is set to be 10, i.e. the redshift when star formation happens.

\section{\label{assumption}Electron-positron pairs excess on 1.4 TeV}

Due to the existence of galactic magnetic field, $e^+e^-$ does not follow a straight line. Their flux after injection from a source can be described by the transport equation,
\begin{equation}
    \frac{\partial n}{\partial t} = D(E)\,\nabla^2\,n+Q_s(E)\delta(\vec{x} - \vec{x}_s)
\end{equation}
where $D(E)$ is the spatial diffusion coefficient. We take $D(E) \simeq 10^{29}\rm{cm}^2/\rm{s}$ for 1.4 TeV $e^+e^-$ propagating in our galaxy. This gives the travel distance of these electrons as $\lambda\sim$O(1) kpc. In order to achieve the peak structure in the observed electron positron flux, the source needs to be nearby the solar system, i.e. $R<0.3$ kpc. In this regime, the stationary solution of the transport equation, neglecting energy loss processes, can be written as
\begin{equation}
    n_e(R, E) = \frac{Q_e(E)}{4\pi\,R\,D(E)}
\end{equation}
The 1.4 TeV peak observed at DAMPE indicates the radial energy density distribution of electrons as $w_e\simeq 1.2\times 10^{-8} \rm{erg}/\rm{cm}^3$ \cite{Yuan:2017ysv}. This translates to the source injection power as, 
\begin{eqnarray}
    \dot Q = &&5\times 10^{32}\rm{erg}/\rm{s}\bigg(\frac{R}{0.1 \rm{kpc}}\bigg) \bigg(\frac{D(E)}{10^{29}\rm{cm}^2/\rm{s}}\bigg)\nonumber\\ 
    &\times&\bigg(\frac{w_e}{1.2 \times 10^{-18}\rm{erg}/\rm{cm}^3}\bigg)
\label{qdot}    
\end{eqnarray}
Now we use source injection power to extract the properties of the nearby UCMH which gives the observed $e^+e^-$ flux.

\section{\label{cal}Fitting UCMH properties with the electron-positron pairs excess}
In this paper, we focus on a benchmark model with $e^\pm$ thermal-averaged annihilation cross section $\langle\sigma v\rangle=3\times10^{-26}~\rm{cm^3~s^{-1}}$. The annihilation rate per dark matter particle is 
\begin{equation}
\frac{\rho_{UCMH}(r)}{m_{\chi}}\times\langle\sigma v\rangle.
\label{ratepp}
\end{equation}
The total annihilation rate in the volume $dV=4\pi r^2dr$ is obtained by multiplying Eq.~\ref{ratepp} by the total number of particles in the volume:
\begin{equation}
\left(\frac{\rho_{UCMH}(r)}{m_{\chi}}\langle\sigma v\rangle\right)\times\left(\frac{\rho_{UCMH}(r)}{2m_{\chi}}dV\right).
\label{rate}
\end{equation}
Note that the factor of 2 in the denominator comes from the fact that there are two particles involved in every annihilation interaction. 
With the piecewise density profile, the energy injection power $\dot Q$ from a UMCH can be written as:
\begin{equation} \label{eq:Qdot}
\dot Q = \frac{\langle\sigma v\rangle}{2m_{\chi}^2}I\times 2 E_0,
\end{equation}
\label{equ}
with
\begin{gather}
I=\int_VdV\rho_{UCMH}^2(r)\notag\\
=\left(\int_0^{r_c}+\int_{r_c}^{R^0_{UCMH}}\right)4\pi r^2\rho_{UCMH}^2(r)dr\notag\\
    =\int_0^{r_c}4\pi r^2\rho_c^2dr+\int_{r_c}^{R^0_{UCMH}}4\pi\rho_c^2r_c^{9/2}r^{-5/2}dr,
\end{gather}
Here $E_0$=1.4 TeV is the proposed energy of each $\rm{e}\pm$ pairs produced by WIMPs' annihilation. 

Combining Eqs.~\ref{rho_zr}, \ref{rz}, \ref{qdot} \& \ref{eq:Qdot}, we can extract the properties of the UCMH, i.e. the mass of the UCMH $M_{UCMH}$, the effective radius of the UCMH $R_{UCMH}$ and the open angle of the UCMH, as a function of its distance to the solar system. For example, if $R=100\rm{pc}$, the nowadays mass of the UCMH, determined at the time of star formation $z=10$, is obtained in terms of solar mass $M_\odot$,   
\begin{equation}
M_{UCMH}^0=M_{UCMH}|_{z=10}=2.5\times10^{33}{\rm{g}}\approx1.3M_\odot.
\end{equation}
If $R=300\rm{pc}$, we get 
\begin{equation}
M_{UCMH}^0=M_{UCMH}|_{z=10}=7.5\times10^{33}{\rm{g}}\approx3.8M_\odot.
\end{equation}

The relation between the distance to the UCMH and the mass of this UCMH is shown in Fig.~1. We see that as the distance to this UCMH increases, the mass required to produce sufficient $\rm{e}\pm$ flux increases accordingly. In Fig.~2, we show how the effective radius of such UCMH scales with its distance to the solar system. Furthermore, given the effective radius and the distance to the UCMH, one can estimate the open angle of this UCMH, which is shown in Fig.~3. The open angle is particularly interesting when correlating the electron positron channel with other channels, such as gamma ray \cite{Yuan:2017ysv} and neutrinos {\bf (cite)}.

It is important to study the fraction of DM stored in the UCMHs. Assuming small density perturbation at O($10^{-5}$) to start with, the numerical simulations indicates a very low probability to generate an over-dense regime in order to explain DAMPE excess at 1.4 TeV \cite{Yuan:2017ysv}.  On the other hand,  in Fig.~4, we show the fraction of DM stored in UCMHs in order to have at least one UCMH at certain distance away from the solar system. We find that the DM fraction in UCMHs can be as small as O($10^{-5}$), which can be easily obtained from small scale density fluctuations. Thus the UCMH scenario provides a natural explanation to the DAMPE excess. Note we set the mean DM density around solar system $\bar\rho_\chi=0.4 {\rm GeV cm^{-3}}$.

\begin{figure}[htbp]
\label{M-r}
\centering
\includegraphics[width=0.95\linewidth]{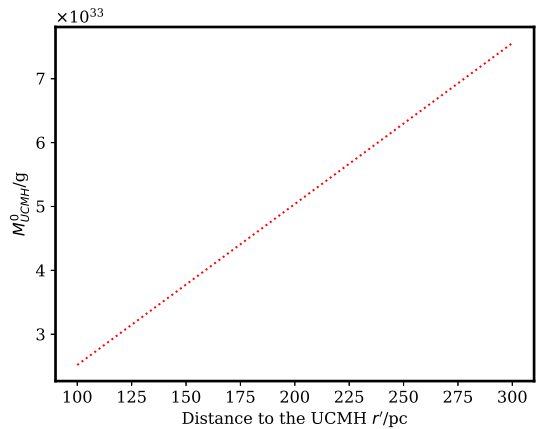}
\caption{The preferred value of UCMH mass as a function of the distance between UCMH and the solar system in order to fit the $e^+e^-$ DAMPE excess.}
\end{figure}

\begin{figure}[htbp]
\label{Radius-R}
\centering
\includegraphics[width=0.95\linewidth]{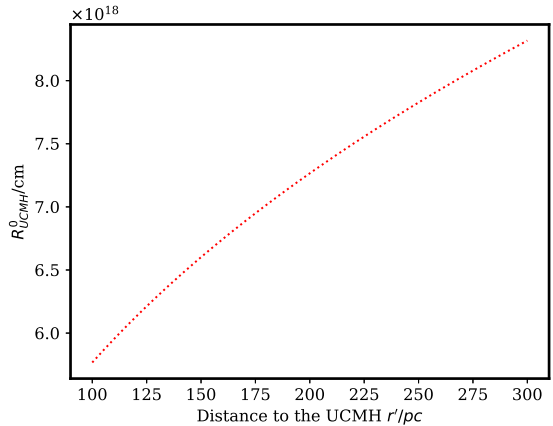}
\caption{The preferred value of UCMH effective radius as a function of the distance between UCMH and the solar system in order to fit the $e^+e^-$ DAMPE excess.}
\end{figure}

\begin{figure}[htbp]
\label{angular_separation}
\centering
\includegraphics[width=0.95\linewidth]{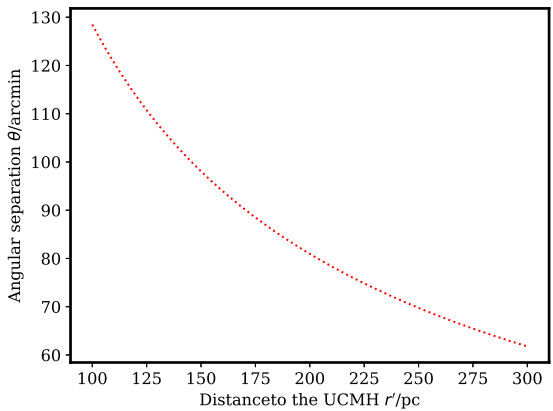}
\caption{The open angle of the UCMH as a function of the distance between UCMH and the solar system.}
\end{figure}

\begin{figure}[htbp]\label{fraction}

\centering
\includegraphics[width=0.95\linewidth]{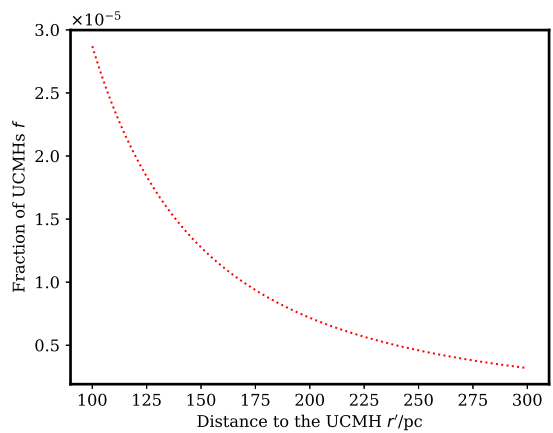}
\caption{The fraction of DM stored in UCMH in order to have at least one UCMH near by the solar system.}
\end{figure}

\section{Discussion}

The tentative peak at 1.4 TeV in $e^+e^-$ spectrum measured by DAMPE motivates the possibility of dark matter annihilation in nearby DM sub-halo. The diffusion of cosmic-ray electrons and positrons can easily smear the sharp peak feature in the energy spectrum across ~kpc propagation scale. Thus the over-dense regime of DM has to be near the solar system. Ordinary collisionless cold dark matter starting with O($10^{-5}$) density fluctuation has a low probability to generate such nearby sub-halo. On the other hand, UCMHs are natural consequences of moderate density perturbations in the early universe. A nearby UCMH may be a good candidate for such dark matter over-dense object in order to explain the $e^+e^-$ excess measured by DAMPE. 

With simple assumptions, such as radial in-fall approximation, the parameters of UCMH profiles are correlated and only few intrinsic parameters remain to be determined, including $\{M_{UCMH}, f_\chi, \langle\sigma v\rangle\}$. Assuming a typical value for thermal-averaged WIMP annihilation cross section, we have determined $M_{UCMH}$ as a function of the distance between the nearby UCMH and the solar system. Furthermore, we find that only a small fraction of DM, i.e. O($10^{-5}$), allocated in UCMHs is enough to induce at least one UCMH nearby in order to explain the DAMPE signal through dark matter annihilation. This can be naturally achieved by moderate density fluctuations at the early time.

\section{Acknowledgement}
YZ thank the support of grant from the Office of Science
and Technology, Shanghai Municipal Government (No.
16DZ2260200). YZ is also supported by US Department
of Energy under grant de-sc0007859.

\bibliography{reference}

\end{document}